\documentclass[fleqn]{2023SCGE}
\begin{document}

\ensubject{subject}

\ArticleType{Article}
\SpecialTopic{}
\Year{}
\Month{}
\Vol{}
\No{}
\DOI{}
\ArtNo{}
\ReceiveDate{}
\AcceptDate{}

\title{Silk damping in scalar-induced gravitational waves: a novel probe for new physics}{Silk damping in scalar-induced gravitational waves: a novel probe for new physics}

\author[1,2]{Yan-Heng Yu}{}%
\author[1]{Sai Wang}{{wangsai@ihep.ac.cn}}

\AuthorMark{Y.-H. Yu}

\AuthorCitation{Y.-H. Yu, S. Wang}

\address[1]{Theoretical Physics Division, Institute of High Energy Physics, Chinese Academy of Sciences, Beijing 100049, China}
\address[2]{School of Physics, University of Chinese Academy of Sciences, Beijing 100049, China}


\abstract{Silk damping is well known in the study of cosmic microwave background (CMB) and accounts for suppression of the angular power spectrum of CMB on large angular multipoles. In this article, we study the effect of Silk damping on the scalar-induced gravitational waves (SIGWs). Resulting from the dissipation of cosmic fluid, the Silk damping notably suppresses the energy-density spectrum of SIGWs on scales comparable to a diffusion scale at the decoupling time of weakly-interacting particles. The effect offers a novel observable for probing the underlying particle interaction, especially for those mediated by heavy gauge bosons beyond the standard model of particles. We anticipate that pulsar timing arrays are sensitive to gauge bosons with mass $\sim10^{3}-10^{4}\,\mathrm{GeV}$, while space- and ground-based interferometers to those with mass $\sim10^7-10^{12}\,\mathrm{GeV}$, leading to essential complements to on-going and future experiments of high-energy physics.}



\maketitle


\begin{multicols}{2}
\section{Introduction}\label{section1}
While the standard model (SM) of particle physics serves as the cornerstone of modern physics, open questions such as neutrino masses, dark matter, and matter-antimatter asymmetry suggest the potential existence of new physics. 
A natural paradigm for constructing theories of beyond the standard model (BSM) involves introducing extended gauge symmetries and assuming them to break at high energy scales, which are often accompanied by the presence of heavy new gauge bosons. 
Detecting these heavy bosons are essential to test the new physics. 
However, if the energy scale far exceeds the TeV range, the particle collider experiments would face challenges in detection. 
The early Universe is an extremely
\Authorfootnote

\noindent 
high-energy laboratory far beyond the reach of these
terrestrial experiments.
As a direct probe to the early Universe, gravitational wave (GWs) of cosmological origins provide a wealth of observables to characterize the processes related to new physics.
With strong evidence for GW background \cite{NANOGrav:2023gor,EPTA:2023fyk,Reardon:2023gzh,Xu:2023wog}, future GW observations would be a significant complementation to particle physics methods for exploring new physics, and any potential imprints should be considered seriously.

Scalar-induced gravitational waves (SIGWs) \cite{Mollerach:2003nq,Ananda:2006af,Baumann:2007zm,Assadullahi:2009jc,Espinosa:2018eve,Kohri:2018awv,Domenech:2021ztg} are an important GW background produced in the early Universe, with close connections to many attractive problems like the inflationary dynamics, the equation-of-state of the early Universe, primordial black holes, etc.
Since they are produced from the nonlinear couplings of linear cosmological scalar perturbations, they could serve as an sensitive indicator of the hydrodynamic properties of the cosmic fluid.
Most of existing studies about SIGWs simply assumed the cosmic plasma as a perfect fluid, which is often a good approximation for the radiation-dominated (RD) era.
However, the dissipative effects in the cosmic plasma should not be simply neglected when the interaction rate of particles is comparable to or smaller than the Hubble expansion rate. 
The diffusion and thermalization of these near-decoupling particles would erase the perturbations in the cosmic plasma inside a typical scale $k_D^{-1}$, causing the damping of the curvature perturbations on the scales smaller than $k_D^{-1}$.
This effect is well-known as ``Silk damping'' particularly for the photon-baryon plasma after big-bang nucleosynthesis, resulting in a depression of the angular power spectrum of cosmic microwave background (CMB) on large angular multipoles, i.e., $\ell \gtrsim 10^3$ \cite{Silk:1967kq}. 
Similarly, the Silk damping of curvature perturbations in the earlier Universe should suppress the energy-density spectrum of SIGWs. 

In this article, we investigate the effect of ``Silk damping" on SIGWs and propose that it could serve as a novel probe for new physics. 
This is manifested as the distinctive frequency-dependent suppression of the SIGW spectrum, arising from the decoupling of certain weakly-interacting particles.
These particles include neutrinos within SM, as well as potential particles in BSM with weaker interactions, mediated by heavier gauge bosons than the $W$ bosons. 
The Silk damping of curvature perturbations is governed by these weak particle interactions, indicating that the corresponding GW imprints could encode crucial information of physics on extremely high energy scales. 
The Silk damping in SIGWs holds the potential for multi-band detection through future GW detectors, including the pulsar timing arrays (PTAs) and space- or ground-based GW interferometers.
Furthermore, these GW observations could complement the on-going and future collider experiments and cosmic-ray studies, offering insights into the construction of new-physics models.

\section{Cosmic plasma as imperfect fluid}\label{sec:2}
We firstly consider the hydrodynamical property of cosmic plasma. 
At the earliest time, all the particles in the cosmic plasma are relativistic and collide frequently, with the energy-momentum tensor being well modeled as the form of perfect fluid $\mathcal{T}_{\alpha\beta}=pg_{\alpha\beta}+(\rho+p)\,u_\alpha u_\beta$, with $p$, $\rho$, and $u_\alpha$ being the pressure, energy density, and velosity four-vector, respectively.
As the temperature $T$ drops down, the mean free time of the particles increases, which may lead to dissipative effects, causing the cosmic plasma to deviate from the perfect fluid approximation.
In the theory of a relativistic imperfect fluid, these deviations are generally parameterized by the shear viscosity $\eta$, bulk viscosity $\xi$, and heat conduction $\chi$, and additional term $\Delta \mathcal{T}_{\alpha\beta}(\eta,\xi,\chi)$ should be incorporated into the total energy-momentum tensor \cite{Eckart:1940te,Weinberg:1971mx}.
This additional term lets the linear perturbations $\delta p$, $\delta \rho$, and $\delta u_\alpha$ to evolve with a decay rate $\Gamma\propto k^2$ \cite{Weinberg:1971mx}, where we introduce $k=|\mathbf{k}|$ and $\mathbf{k}$ is the Fourier mode of the perturbations.

In the RD era, the evolution of scalar perturbations $\phi_\mathbf{k}$ can be modified by the dissipative effects in the cosmic plasma.
We connect $\phi_\mathbf{k}$ to the primordial curvature perturbations $\zeta_\mathbf{k}$ through $\phi_\mathbf{k}(\tau)=(2/3)\,\Phi(k,\tau)\,\zeta_\mathbf{k}$, where $\Phi(k,\tau)$ is the scalar transfer function at the conformal time $\tau$.
Considering $-k^2\phi_\mathbf{k}\propto \delta\rho$, we find that $\Phi(k,\tau)$ has an additional exponential suppression due to the decay of $\delta \rho$, namely,
\begin{equation}\label{eq:trans}
    \Phi(k,\tau)\simeq
    \frac{9}{(k \tau)^2}
    \left(
    \frac{\sqrt{3}}{k \tau}\sin{\frac{k \tau}{\sqrt{3}}}
    -\cos{\frac{k \tau}{\sqrt{3}}}
    \right)
    \times 
    e^{-k^2/k_D^2(\tau)}\ ,
\end{equation}
where $k_D^{-1}$ stands for the typical Silk damping scale. 
For $k\gtrsim k_D$, the factor of $ \mathrm{exp}(-k^2/k_D^2)$ dominates the decay, otherwise if $k\ll k_D$, Eq.~(\ref{eq:trans}) gives the standard result without dissipation. 

\section{Microscopic origin of Silk damping}\label{sec:3}
The Silk damping scale $k_D^{-1}$ is determined by the interaction of particles in cosmic plasma.
In the RD Universe, the effects from the bulk viscosity and heat conduction can be neglected \cite{Weinberg:1971mx}, and thereby $k_D$ is approximately given by the shear viscosity, namely,
\begin{equation}\label{eq:kD}
    k_D^{-2}(\tau)\simeq 
    \int_0^\tau \, \mathrm{d} \tau'\ 
    \frac{2\,\eta(\tau')}{3\,a(\tau')\,[\rho(\tau')+p(\tau')]}\ .
\end{equation}
In Eq.~(\ref{eq:kD}), $a$ is a scale factor of the Universe, and $\eta$ is contributed by photons, neutrinos, and underlying weakly-interacting particles of BSM, denoted as $X$, given by \cite{Weinberg:1971mx,10.1093/mnras/202.4.1169,Jeong:2014gna}
\begin{equation}\label{eq:shear viscosity}
   \eta(\tau)\simeq
   \frac{16}{45}\rho_\gamma t_\gamma 
   +\sum_{j=\nu,X,\, \ldots}\,\frac{4}{15}\rho_j t_j \,\Theta(\tau_{j,\mathrm{dec}}-\tau)\ ,
\end{equation}
where $\tau_{j,\mathrm{dec}}$ is the decoupling time, and $t_j=1/(n_j \langle\sigma_j v \rangle)$ denotes the mean free time with $n_j$ being the number density of the target particles and $\langle\sigma_j v \rangle$ the thermally-averaged cross section.

\begin{figure}[H]
\includegraphics[scale=0.65]{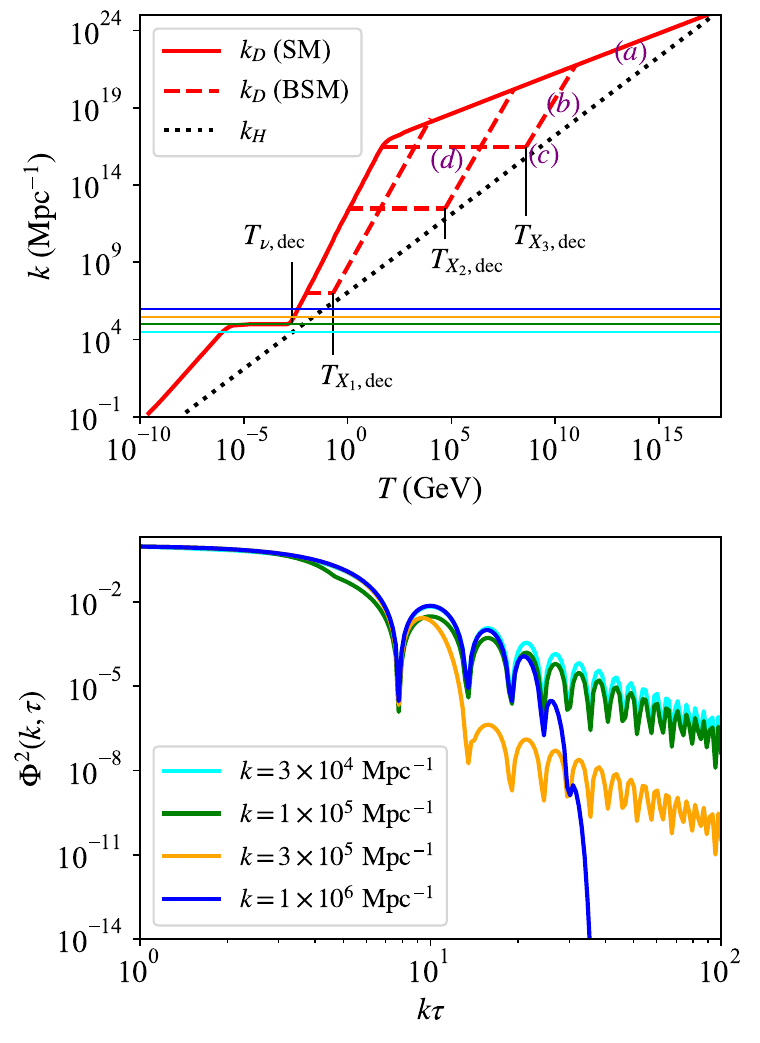}
\caption{ Upper panel: Silk damping parameter $k_{D}$ versus the cosmic temperature $T$. Red solid line stands for SM \cite{Jeong:2014gna}, while red dashed lines denote BSM. We plot the comoving Hubble horizon $k_H=aH$ for comparison. Lower panel: $\Phi^2(k,\tau)$ versus $k\tau$ when considering $k/ \mathrm{Mpc}^{-1}=3\times 10^4 $ (cyan), $10^5$ (green), $3\times 10^5$ (orange), and $10^6$ (blue), which are also marked as the corresponding colors in the upper panel.} 
\label{figure1}
\end{figure}

The evolution of $k_D$ contains crucial information about the thermal history of the early Universe. 
According to Eqs.~(\ref{eq:kD}-\ref{eq:shear viscosity}), the particle species dominating $k_D$ should be not only abundant but also weakest-interacting (but are still coupled with the cosmic plasma).
Within the SM, $k_D$ is controlled by the weak interaction of neutrinos before their decoupling at $T_{\nu,\mathrm{dec}}\simeq 1.5\, \mathrm{MeV}$.
After $T\lesssim 2\, \mathrm{keV}$, photons dominate the evolution of $k_D$ through their Thompson scattering with electrons \cite{Jeong:2014gna}. 
Furthermore, if there exist a new kind of abundant particles, denoted as $X$, whose interaction is mediated by a gauge boson with a larger mass $M'$ than that of $W$ boson, i.e., $M'> M_W \sim 100\,\mathrm{GeV}$, they should be feebler-interacting and decouple earlier than neutrinos, dominating the Silk damping before their decoupling. 
In Figure~\ref{figure1}, we plot the temperature-dependence of $k_D$ in the RD era, considering the contributions from photons, neutrinos, and $X$ particles. 

We focus on Silk damping at $T > 2\, \mathrm{keV}$, where the dominating weakly-interacting particles are neutrinos or $X$.
The temperature-dependence of $k_D$ can be summarized as the following four processes, as marked in the upper panel of Figure~\ref{figure1}.
(a) In the earliest era $T\gg M$ ($M=M_W$ for neutrinos and $M=M'$ for $X$ particles), the gauge boson is considered as massless and the cross section is roughly given by $\langle\sigma_j v \rangle\sim \alpha^2/T^2$ \cite{Kolb:1990vq}, where $\alpha=g^2/(4\pi)$ and $g$ is assumed as a weak coupling strength in this work. 
Therefore, the $k_D$ behaves as $k_D\propto T^{1/2}$, obtained by Eqs.~(\ref{eq:kD}-\ref{eq:shear viscosity}). 
(b) Once the temperature decreases to $T\lesssim M$, the massive gauge boson gives $\langle\sigma_j v \rangle\sim G^2 T^2$ with $G=\alpha/M^2$ \cite{Kolb:1990vq}, leading to $k_D\propto T^{5/2}$.
(c) When the weakly-interacting particles decouple, i.e., $t_j\sim H^{-1}$ with $H$ being the Hubble parameter at decouping time, Figure~\ref{figure1} illustrates that the Silk damping scale $k^{-1}_{D,j}\equiv k_D^{-1}(\tau_{j,\mathrm{dec}})$ would be smaller but not much smaller than the comoving Hubble horizon $k_{H,j}^{-1}\equiv k_H^{-1}(\tau_{j,\mathrm{dec}})$.
\footnote{A rough estimate can be derived as follows, neglecting the change in the effective number of relativistic degrees of freedom. 
Eqs.~(\ref{eq:kD}-\ref{eq:shear viscosity}) give $k_{D,j}^{-2}
    \sim \int_0^{\tau_{j,\mathrm{dec}}} \mathrm{d} \tau\ (2\rho_j t_j)/(15 a\rho)
    \sim \int_\infty^{T_{j,\mathrm{dec}}} \mathrm{d} T\ (-aHT)^{-1} (2\rho_j t_j)/(15 a\rho)\, .
    $
    As the integrand is proportional to $T^{-6}$, we have 
    $k_{D,j}^{-2} \sim (2\rho_j t_j)/( 75a^2H\rho)\big|_{T=T_{j,\mathrm{dec}}}
    \sim (2\rho_j) / (75a^2 H^2\rho)
    \sim (2\rho_j)/(75\rho)\times k^{-2}_{H,j}\ ,$ where the decoupling relation $t_j\sim H^{-1}$ is used in the second step.
    Hence, we obtain $k_{D,j}\sim \sqrt{(75\rho)/(2\rho_j)}\  k_{H,j}$, implying that $k_D$ is comparable to $k_H$ at the decoupling time of abundant weakly-interacting particles. 
    See also the numerical results in Ref.~\cite{Jeong:2014gna}. }
As shown below, this general property would significantly affect the production of SIGWs.
(d) The weakly-interacting particles no longer contribute to Silk damping after their decoupling, and thereby $k_D$ keeps a constant until another particle species dominates $k_D$ \cite{Jeong:2014gna}. 
Here, we should note that the evolution of $k_D$ above the electroweak scale may contain unknown uncertainties and is determined by the specific new-physics model we consider.
However, this issue would not change our overall picture.

\section{Energy-density spectrum of SIGWs}\label{sec:4}
\begin{figure}[H]
\includegraphics[scale=0.65]{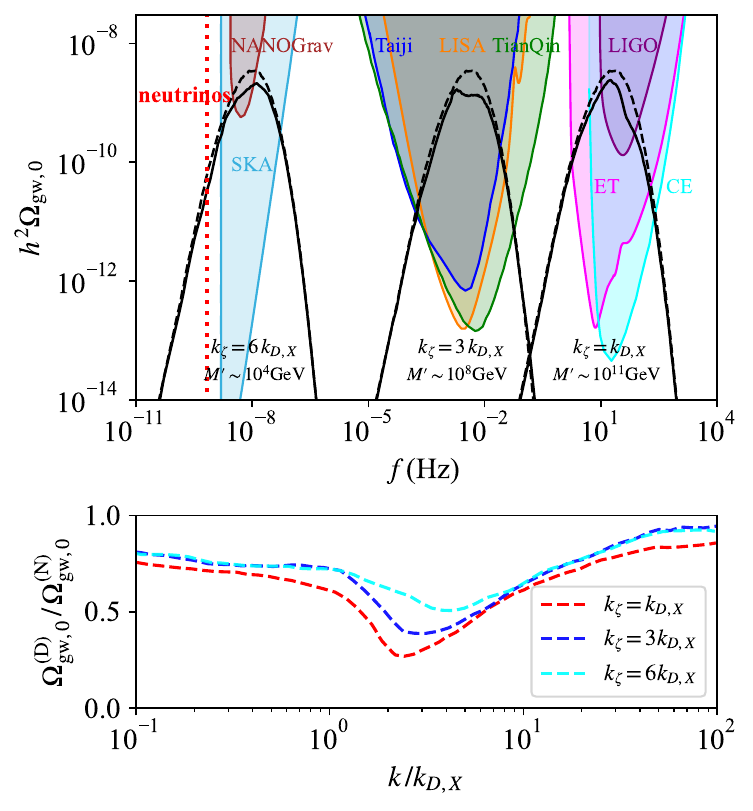}
\caption{ Upper panel: Present-day physical energy-density fraction spectra $h^2 \Omega_\mathrm{gw,0}$ with (black solid lines) and without (black dashed lines) the Silk damping. From left to right, $k_\zeta$ and $k_{D,X}$ are assumed as $k_\zeta=6\, k_{D,X}=6\times10^{7}\, \mathrm{Mpc}^{-1}$, $k_\zeta=3\,k_{D,X}=9\times 10^{12}\, \mathrm{Mpc}^{-1}$, and $k_\zeta=k_{D,X}=3\times10^{16}\, \mathrm{Mpc}^{-1}$, respectively. Other parameters are set as $ \mathcal{A}_{\zeta}=3\times10^{-2}$ and $\sigma=1$. The shaded regions show the sensitivities of NANOGrav (brown), SKA (light blue), LISA (orange), Taiji (blue), TianQin (green), LIGO (purple), ET (pink), and CE (cyan). A vertical line (red dotted) marks the frequency of the maximal suppression of $\Omega_\mathrm{gw,0}$ due to the neutrino decoupling. Lower panel: The ratio ${\Omega}_{\mathrm{gw},0}^{\mathrm{(D)}}\,/\, {\Omega}_{\mathrm{gw},0}^{\mathrm{(N)}}$ versus $k/k_{D,X}$ with $k_\zeta=k_{D,X}$ (red dashed line), $k_\zeta=3\,k_{D,X}$ (blue dashed line), and $k_\zeta=6\,k_{D,X}$ (cyan dashed line).} 
\label{figure2}
\end{figure}

We study the effect of Silk damping on the production of SIGWs. 
We adopt an enhanced primordial curvature power spectrum on small scales, parametrized as \cite{Balaji:2022dbi}
\begin{equation}\label{eq:Pzeta}
    \mathcal{P}_{\zeta}(k)
    =\frac{\mathcal{A}_{\zeta}}{\sqrt{2\pi}\sigma}\  e^{-\frac{1}{2 \sigma^2}\ln^{2} (k/k_\zeta)}\ , 
\end{equation}
where $\sigma$ represents the spectral width and $\mathcal{A}_\zeta$ the spectral amplitude at the peak wavenumber $k_{\zeta}$.
The energy-density fraction spectrum of SIGWs in the RD era is given by \cite{Espinosa:2018eve,Kohri:2018awv}
\begin{equation}\label{eq:ogw}
\begin{aligned}
    &\Omega_\mathrm{gw}(k,\tau)
    =\ 
    \frac{k^2}{6\,a^2H^2}
    \int _0^{\infty} \mathrm{d} u
    \int _{\lvert 1-u \rvert} ^{\lvert 1+u \rvert} \mathrm{d} v
    \ 
    \mathcal{P}_{\zeta}(uk)
    \mathcal{P}_{\zeta}(vk)
    \\
    &\quad\quad\quad\quad
    \times
    \bigg[\frac{4 v^2-\left(1+v^2-u^2\right)^2}{4uv}\bigg]^2\
    \overline{I^{2}(u,v,k,\tau)}\ .
\end{aligned}
\end{equation}
Here, the overbar represents an oscillation average, and the kernel function $I(u,v,k,\tau)$ is defined as
\begin{equation}\label{eq:kernel}
\begin{aligned}
    I(u,v,k,\tau)
    =\int _0^{\tau}\mathrm{d}\overline{\tau}\, 
    \frac{\overline{\tau}}{\tau}
    \, k\sin{(k\tau-k\overline{\tau})}
    \, f(u,v, k,\overline{\tau})\ ,
\end{aligned}
\end{equation}
where $f(u,v,k,\tau)$ encodes the evolution information of the source terms of SIGWs, namely,
\begin{equation}\label{eq:f_phi}
\begin{aligned}
   &f(u,v,k,\tau)
   =\frac{4}{3}
   \Phi(uk,\tau) \Phi(vk,\tau)
   +\frac{4}{9} 
   \tau^2\, 
   \dot{\Phi} (uk,\tau)
   \dot{\Phi} (vk,\tau)
   \\
   &\quad\quad\quad 
   +\frac{4}{9}
   \Big[
   \tau \, \Phi (uk,\tau)
   \dot{\Phi} (vk,\tau)
   +\tau\,  \dot{\Phi} (uk,\tau)
   \Phi (vk,\tau)
   \Big]
   \ .
\end{aligned}
\end{equation}
Here, an overdot denotes a derivation of the variable $\tau$, and $\Phi(k,\tau)$ can be obtained by Eqs.~(\ref{eq:trans}-\ref{eq:shear viscosity}).
In the upper panel of Figure~\ref{figure2}, we depict the present-day physical energy-density fraction spectra of SIGWs, namely,
\begin{equation}\label{eq:ogw0}
   h^2{\Omega}_{\mathrm{gw},0} (f) \simeq h^2\Omega_{\mathrm{rad}, 0}\, {\Omega}_\mathrm{gw} (k,\tau)\big|_{k=2\pi f}\ ,
\end{equation}
where $h^2\Omega_{\mathrm{rad},0}=4.2\times 10^{-5}$ is the present-day physical energy-density fraction of radiations \cite{Planck:2018vyg}.
The black solid lines, labeled as $h^2{\Omega}_{\mathrm{gw},0}^{\mathrm{(D)}}$, consider the Silk damping caused by the decoupling of $X$ (corresponding to the red dashed lines in the upper panel of Figure~\ref{figure1}), while the black dashed lines, labeled as $h^2{\Omega}_{\mathrm{gw},0}^{\mathrm{(N)}}$, do not incorporate this effect.
The ratio of the former to the latter is also presented in the lower panel of Figure~\ref{figure2}.

Figure~\ref{figure2} demonstrates that Silk damping notably suppresses the SIGW spectrum for modes $k \sim (2-5)\times k_{D,j}$, with the ratio ${\Omega}_{\mathrm{gw},0}^{\mathrm{(D)}}\,/\, {\Omega}_{\mathrm{gw},0}^{\mathrm{(N)}}\sim 0.3-0.5$. 
These results can be explained as follows. 
For the scalar perturbation $\phi_\mathbf{k}$, the period of the efficient SIGW production begins when $\phi_\mathbf{k}$ reenters the horizon, i.e., $k=k_H$, and almost ends when the Silk damping becomes important, i.e., $k=k_D$. 
According to the upper panel of Figure~\ref{figure1}, this period should be the shortest for $k$ slightly larger than $k_{D,j}$, since $k_D$ and $k_H$ becomes very close only near the decoupling times of weakly-interacting particles.  
Therefore, a notable suppression in the SIGW spectrum is expected at $k\sim k_{D,j}$. 
As a visual illustration, the lower panel of Figure~\ref{figure1} shows that for $k$ slightly larger than $k_{D,\nu}$ (blue and orange lines), $\phi_\mathbf{k}$ undergoes exponential damping shortly after it begins to oscillate inside the horizon, leading to reduced SIGW production. 
Conversely, for other modes (cyan and green lines), the damping effects can be disregarded.
The analysis above is also consistent with our picture that the dissipative effects should be pronounced near the decoupling times of weakly-interacting particles.

The scale-dependent suppression on SIGW spectrum due to Silk damping has significant implications in cosmology and particle physics. 
This featured suppression on the scales $k\sim k_{D,j}$ approximately tells us the decoupling times of weakly-interacting particles, which encodes information about the thermal history of the early Universe. 
It also provides new observables to probe these underlying particle interactions, as will be discussed below.

\subsection{Neutrinos}
In the SM content, the Silk damping in SIGWs mainly stems from the neutrinos, which decouple at $T_{\nu,\mathrm{dec}}\simeq 1.5\, \mathrm{MeV}$ with $k_{D,\nu}\simeq 10^5\, \mathrm{Mpc}^{-1}$.
As a result, $\Omega_\mathrm{gw,0}(f)$ is suppressed around nanohertz frequencies, which are near the lowest observable frequency range of the PTA experiments, say, $f_\mathrm{min}\simeq 2\, \mathrm{nHz}$. 
The obvious suppression occurs around $f\sim (0.5-1)\,  \mathrm{nHz}$, as marked in the upper panel of Figure~\ref{figure2}.
With increase of observational duration $t_\mathrm{obs}$, PTA experiments, including the North American Nanohertz Observatory for Gravitational Waves (NANOGrav) \cite{Jenet:2009hk}, the European Pulsar Timing Array (EPTA) \cite{Kramer:2013kea}, the Parkes Pulsar Timing Array (PPTA) \cite{Manchester:2012za}, the Chinese Pulsar Timing Array (CPTA) \cite{2016ASPC..502...19L}, and the Square Kilometer Array (SKA) \cite{2009IEEEP..97.1482D}, are projected to detect GWs with lower frequencies, i.e., $f_\mathrm{min}\sim 1/t_\mathrm{obs}$  \cite{Maggiore:2018sht}, enabling a more comprehensive investigation into the Silk damping in SIGWs caused by neutrino decoupling.
If the GW background reported by the PTA collaborations \cite{NANOGrav:2023gor,EPTA:2023fyk,Reardon:2023gzh,Xu:2023wog} is mainly attributed to SIGWs, the slope of $\Omega_\mathrm{gw,0}(f)$ at $f\gtrsim 1\,$nHz could be enlarged compared with that without considering Silk damping, which may impact data analysis in the future.
In the coming decades, this damping has chances to serve as a distinctive feature of SIGWs in comparison to other potential sources of nanohertz GWs like supermassive black hole binaries.

\subsection{New physics}
Additional gauge symmetry breaking at high energy scale is a feature of many extensions of the SM, and detecting the possible associated charged (or neutral) gauge boson $W'$ (or $Z'$) is of great importance to search for new physics.
However, the mass of the extra gauge boson, $M'$, is considered as an independent parameter in many BSM models, while conventional particle-physics methods may have some limitations to cover the parameter space of $M'$ much above TeV.
As a novel approach, the Silk damping in SIGWs provide a general framework to scan the parameter $M'$ in a wider range.
The weakly-interacting particles mediated by some gauge bosons with $M' > 100\, \mathrm{GeV}$ decouple earlier than neutrinos with $\left(k_D/1\,\mathrm{Mpc}^{-1}\right) \sim \mathrm{few}\times 10^2\,\left(M'/1\,\mathrm{GeV}\right)^{4/3}$, and are expected to suppress $\Omega_\mathrm{gw,0}(f)$ at frequencies
\begin{equation}\label{eq:fM'}
    \left(\frac{f}{1\,\mathrm{Hz}}\right)\sim \mathrm{few}\times10^{-13}\,\left(\frac{M'}{1\, \mathrm{GeV}}\right)^{4/3}\ .
\end{equation}
This featured suppression is potentially detectable for PTA experiments and GW interferometers in the future, offering valuable implications to the model-building works for new physics.
 
The PTA experiments have the capability to detect the Silk damping in SIGWs with a mass of typically $M'\sim \mathcal{O}(10^3-10^4)\, \mathrm{GeV}$. 
Numerous extensions of the SM involve the introduction of new gauge bosons with a TeV mass, including the Littlest Higgs \cite{Arkani-Hamed:2001nha,Arkani-Hamed:2002ikv}, the sequential standard model \cite{Altarelli:1989ff}, some supersymmetric grand unified theories (GUT) \cite{Cvetic:1995rj,Cvetic:1996mf,Lykken:1996kz,Espinosa:1997ji}, among others. 
These new physics models at TeV scale is of particular interest because of its potential detectability in next-generation colliders. 
The GW signals would provide a crucial crosscheck with particle physics experiments in the Large Hadron Collider (LHC) and future colliders such as the Future Circle Collider (FCC) \cite{FCC:2018evy}, the International Linear Collider (ILC) \cite{ILCInternationalDevelopmentTeam:2022izu}, and the Circular Electron Positron Collider (CEPC)  \cite{CEPCStudyGroup:2018rmc,CEPCStudyGroup:2018ghi}.
These integrated studies will enhance our understanding of TeV-scale physics.
In the upper panel of Figure~\ref{figure2}, the solid black line on the left represents an example of $M'\sim 10^4\,\mathrm{GeV}$, illustrating the potential detectability for PTA experiments.

In the higher-frequency region, a range of GW interferometers are able to cover frequencies in the vicinity of $\mathcal{O}(10^{-4}-10^3)\, \mathrm{Hz}$, corresponding to the parameter space of $M'$ around $\mathcal{O}(10^{7}-10^{12})\, \mathrm{GeV}$. 
These scales are considerably beyond the reach of terrestrial experiments, yet exploring them may be essential for certain new physics models.
For instance, all GUT featuring gauge groups larger than $SU(5)$ anticipate the existence of extra gauge bosons, but their masses are generally unconstrained by the theories, allowing $M'$ to widely span from the electroweak scale to the GUT scale \cite{Leike:1998wr}. 
The Silk damping in SIGWs also presents an opportunity for a combined exploration of heavy bosons with high-energy cosmic ray observations, such as those conducted by the the Pierre Auger Observatory (PAO)  \cite{PierreAuger:2015eyc}, the Telescope Array (TA) \cite{TelescopeArray:2008toq}, the Large High Altitude Air Shower Observatory (LHAASO) \cite{LHAASO:2019qtb}, etc.  
The investigation of the interactions mediated by such heavy gauge bosons may also provide insights into addressing issues such as dark matter.
In the upper panel of Figure~\ref{figure2}, we depict the black solid line on the middle (or right) to illustrate the potential for detecting $M'\sim 10^8\, \mathrm{GeV}$ (or $10^{11}\, \mathrm{GeV}$) in space-based experiments such as  Laser Interferometer Space Antenna (LISA) \cite{LISA:2017pwj,2019BAAS...51g..77T,LISACosmologyWorkingGroup:2022jok}, Taiji \cite{Hu:2017mde}, and TianQin \cite{TianQin:2015yph} (or in ground-based experiments like Advanced LIGO, Virgo, and KAGRA (LVK) \cite{Harry_2010,VIRGO:2014yos,Somiya:2011np}, Einstein Telescope (ET) \cite{Punturo:2010zz} and Cosmic Explorer (CE) \cite{Reitze:2019iox}). 

\section{Conclusion and discussion}\label{sec:5}
In this article, we presented that the dissipation in the cosmic plasma leads to Silk damping of cosmological scalar perturbations, consequently suppressing the production of SIGWs.
The most significant suppression occurs at scales slightly smaller than the diffusion scale, potentially reducing the SIGW spectrum by $\sim 70\%$.
This effect offers novel observables for probing the weak particle interactions, which may be detectable by PTA experiments or GW interferometers in the future.
The measurement of the Silk damping in SIGWs holds significant implications for the search of new physics, particularly those related to remarkably high energy scales that are challenging to investigate in terrestrial experiments.

Compared to the anisotropic stress induced by free-streaming neutrinos or relativistic weakly-interacting particles, which reduces the SIGW spectrum by a factor of at most $\mathcal{O}(0.1)$ on the scales $k\lesssim k_H(\tau_{j,\mathrm{dec}})$ \cite{Bartolo:2010qu,Mangilli:2008bw,Saga:2014jca,Zhang:2022dgx}, Silk damping leads to a more significant suppression of the SIGW spectrum and occurs on smaller scales $k\gtrsim k_H(\tau_{j,\mathrm{dec}})$.
For future GW observations, the dissipative effects are an essential factor to consider.

We plotted Figure~\ref{figure2} under the condition of $k_\zeta$ being close to $k_{D,X}$, where the Silk damping effect is the most pronounced. 
When these two scales do not coincide, such as for $k_\zeta/k_{D,X} \sim 10$ (or $\sim 50$), Silk damping could still reduce the SIGW spectrum by $\sim40\%$ (or $\sim20\%$). 
This suggests that Silk damping remains an important effect in more general cases.

In our work, we assumed that $X$ and neutrinos have the same abundance at their respective decoupling times. 
In principle, a larger (or smaller) abundance of weakly-interacting particles could increase (or decrease) the suppression on SIGW spectrum, which may provide suggestions about whether these particles can be a leading component of dark matter. 

The origin of the features on $\Omega_\mathrm{gw}$ is always of great interest.
Besides the exotic shape of $\mathcal{P}_\zeta$, which has received extensive attention (e.g., see Ref.~\cite{Braglia:2020taf} and the references therein), our study of Silk damping develops a new territory to reveal the physics underlying these features.
Initiated by our work, future research of Silk damping in other observables, such as the anisotropies of SIGWs \cite{Bartolo:2019zvb,LISACosmologyWorkingGroup:2022kbp,Li:2023qua,Wang:2023ost,Li:2023xtl,Yu:2023jrs,Li:2024zwx} and the mass function of primordial black holes (see reviews, e.g., in Ref.~\cite{Carr:2020gox}), is expected to break degeneracies between the two origins, enhancing our understanding of the early Universe.

Furthermore, similar analysis for the damping effects is applicable to the GWs of other cosmological origins, e.g., the GWs from cosmological phase transition \cite{Guo:2023koq}. 
The imprints of cosmic dissipation on GWs could provide a general approach to probe the particle interactions in the early Universe.

\Acknowledgements{We appreciate Huai-Ke Guo, Jun-Peng Li, Bin Yan, Jing-Zhi Zhou, and Qing-Hua Zhu for helpful discussion. 
This work is supported by the National Natural Science Foundation of China (Grant No. 12175243), the National Key R\&D Program of China No. 2023YFC2206403, the Science Research Grants from the China Manned Space Project with No. CMS-CSST-2021-B01, and the Key Research Program of the Chinese Academy of Sciences (Grant No. XDPB15).}

\InterestConflict{The authors declare that they have no conflict of interest.}


\bibliographystyle{scpma}

\bibliography{ref} 

\end{multicols}
\end{document}